\documentclass[twocolumn,preprintnumbers,amsmath,amssymb]{revtex4}

\usepackage{graphicx}
\usepackage{dcolumn}
\usepackage{bm}
\usepackage[utf8x]{inputenc}
\usepackage{float}

\begin{document}


\title{Mass Spectrum of Exotic X(5568) State via Artificial Neural Network}

\author{Halil Mutuk}
 \email{halilmutuk@gmail.com}
\affiliation{Physics Department, Faculty of Arts and Sciences, Ondokuz Mayis University, 55139, Samsun, Turkey
}%


\begin{abstract}
In this paper, we assume $X(5568)$ exist and study mass spectrum of $X(5568)$ resonance and its hypothetical charmed partner, $X_c$, by Artificial Neural Network method. The obtained
predictions are compared with the experimental data and results of other theoretical works. 
\end{abstract}

\maketitle

\section{\label{sec:level1}Introduction}
With the discovery of the exotic states, i.e., states that cannot be interpreted by the conventional meson (quark-antiquark) and baryon (three quark) pictures, hadron spectroscopy presented some puzzles in the understanding of our nature. 

The milestone of exotic states was the observation of charmoniumlike resonance $X(3872)$ by Belle collaboration in 2003 \cite{1}. It was later confirmed by D0 \cite{2}, CDF II \cite{3} and BABAR \cite{4} collaborations. This observation opened a new era in understanding of our nature and as well as Standard Model framework. Many new charmoniumlike states were observed and this new particle zoo is named $XYZ$ particles. 

The lack of conventional picture of mesons to interpret the underlying structure of the exotic states  paved the way for new theoretical approaches  \cite{5,6,7,8,9}. These states are out of conventional meson picture ($q\bar{q}$). Although there is no consensus about the internal structure of exotic states, some properties of these states are accommodated by tetraquark models, molecular models or updated potential models. Among these approaches, tetraquark model and molecular model got more attention. In tetraquark model, two heavy and two light quarks come in together or mix. These quarks may cluster into the colored diquark-antidiquark, ($Qq$)$+$($\bar{Q}\bar{q}$) doublet. In the molecular model, the exotic particle is thought to be a bound state of two color-singlet mesons, ($Q\bar{q}$) $+$($\bar{Q}q$).

In 2016, a resonance named $X(5568)$ was reported by the D0 collaboration at the $B_s\pi^{\pm}$ invariant mass spectrum with the mass and width, respectively \cite{10}

\begin{eqnarray*}
M&=&5567.8 \pm 2.9 ^{+0.9}_{-1.9} ~ \text{MeV}, \\
\Gamma &=& 21.9 \pm 6.4 ^{+5.0}_{-2.5} ~ \text{MeV}.
\end{eqnarray*}

As it is mentioned in \cite{10}, it is the first observation of a hadronic state with four different flavor quarks. The reason for that is that, as the decay rate of $X(5568) \to B_s \pi^{\pm}$ is much larger than the weak interaction prediction, one can conclude that strong interaction is responsible for this decay. Since strong interactions do not touch the flavor, in the final state there are four different quarks of $B_s=\bar{b}s$ and $\pi^+=u\bar{d}$ which are cannot be created by the vacuum \cite{11}. 

Beside other puzzling features, $X(5568)$ was not observed in $X \to B_s \pi^{\pm}$ channel as reported by LHCb collaboration \cite{12}, the CMS collaboration \cite{13}, the CDF collaboration at Fermilab \cite{14} and ATLAS collaboration of LHC \cite{15}. In 2018, the D0 collaboration announced that they had confirmed the existence of $X(5568)$ from the decay $X(5568) \to B_s \pi^{\pm}$ via a sequent semileptonic decay $B_s^0 \to \mu^{\pm} D_s^{\mp}$ \cite{16} and the results were in consistent except the widht is shifted to
\begin{equation*}
\Gamma= 18.6^{+7.9}_{-6.1} (\text{stat}) ^{+3.5}_{-3.8} (\text{syst}) ~ \text{MeV}.
\end{equation*}
The clear discrepancy between D0 collaboration results and other experimental groups fired a dispute. Since $X(5568)$ may be the first observed exotic state with four different flavors, both theoretical and experimental studies on it can enlighten our realization of quark model. There are different approaches related to $X(5568)$ which calculate masses, widths, decay constants, decay channels and argue about internal structure \cite{17,18,19,20,21,22,23,24,25,26,27,28,29,30,31,32,33,34,35,36,37,38,39,40}. All these studies conclude the mysterious and curious case of $X(5568)$ resonance. 

The hypothetical charmed partner of $X(5568)$, which we will denote it as $X_c$, is composed of $c$,$s$,$u$, $d$ quarks. The related channels for this resonance to be observed can be $X_c \to D_s^- \pi^+$ and $X_c \to D^0 K^0$. The mass, decay constant and widths of this hypothetical resonance was studied in \cite{33}.

In the present study, we adopt diquark-antidiquark and molecular pictures of $X(5568)$ and $X_c$ and calculate mass spectra by artificial neural network for the first time. Artificial neural networks (ANNs) are being used since two decades to solve both ordinary and partial differential equations. They maintain many attractive features compared to known existing semi-analytical and numerical techniques. One of the main advantage of ANNs to solve differential equations is that they require less number of model parameters than any other technique. Besides that, machine learning which is nowadays a hot topic in physics is provided via using ANNs. 

The outline of paper is as follows. In Section \ref{sec:level2}, we introduce ANN formalism and the necessary details for application to quantum mechanics. In Section \ref{sec:level3}, we give our results and discuss. In Section \ref{sec:level4} we summarize our findings. 

\section{\label{sec:level2}Formalism of Artificial Neural Network}
Artificial neural networks are computer systems which are capable of deriving and creating new information and also discovering them through learning which is one of the features of human brain. Neural networks are mimicking versions of biological nervous systems. They are parallel and distributed information processing elements. These elements have their own memory and are connected to each other via weighted connections. 

\begin{figure}[H]
\includegraphics[width=3.4in]{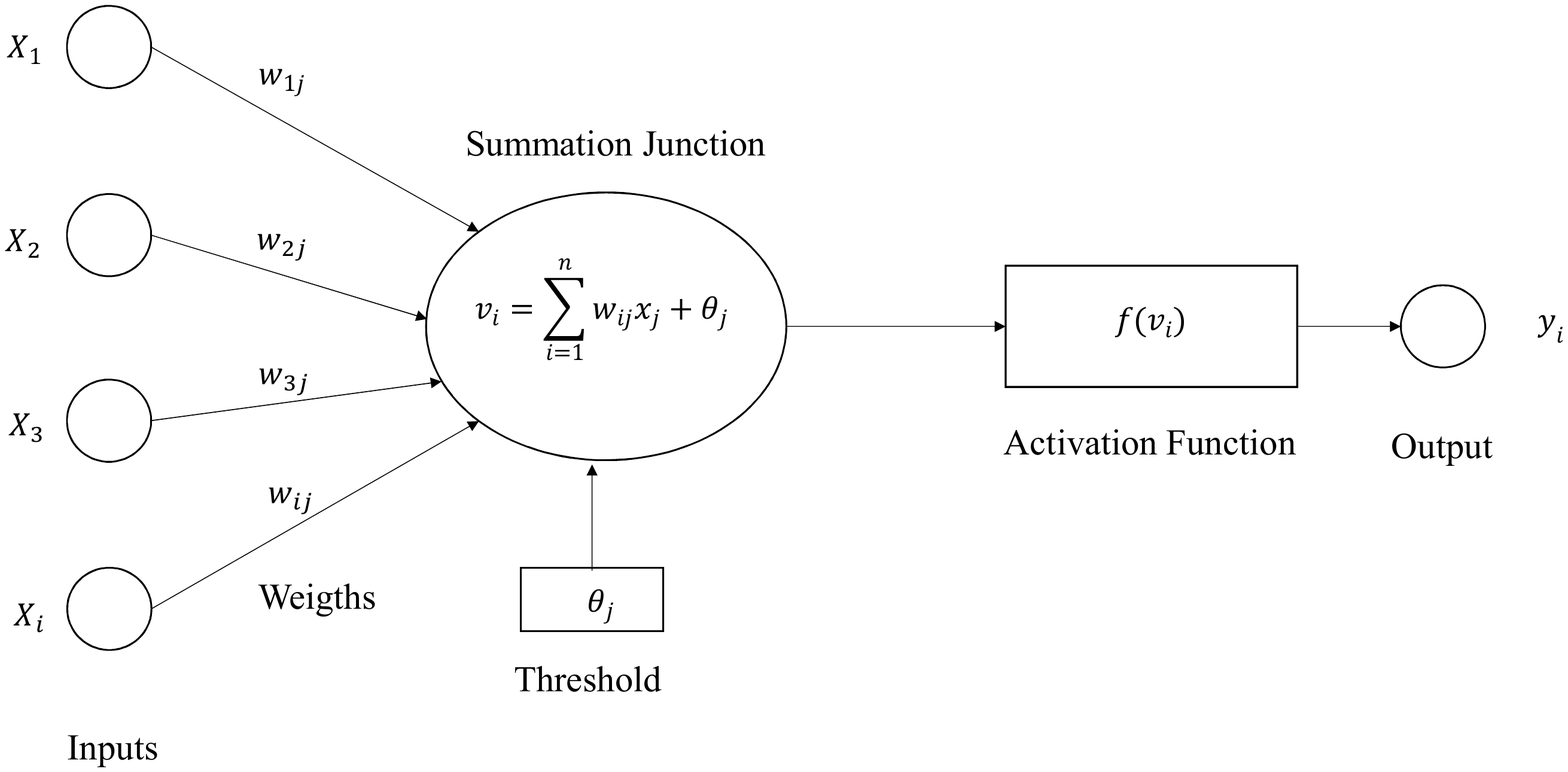}
\caption{\label{fig:1} A model of single neuron}
\end{figure}

\begin{figure}[H]
\includegraphics[width=3.4in]{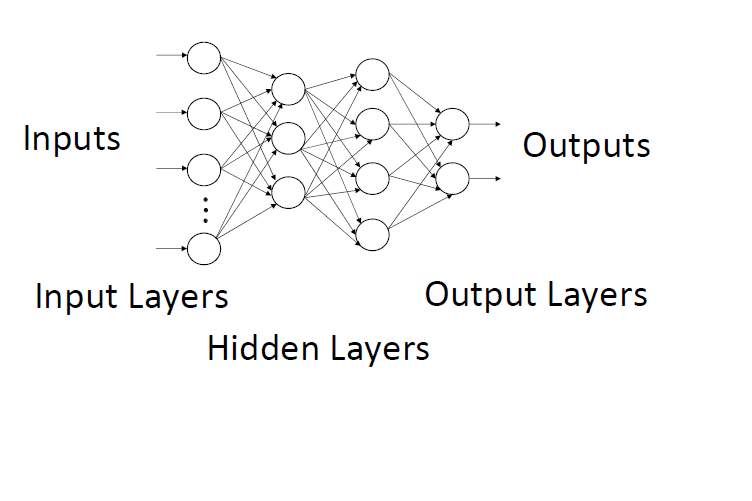}
\caption{\label{fig:2} Multilayer neural network}
\end{figure}

The fundamental ingredient of an artificial neural network is neuron (perceptron in computerized systems) and it is the processing element in a neural network.  Figure \ref{fig:1} represents a single artificial neuron.
Each neuron receives any number of input and produces only one output. If this output comes from input layers, it will be an input for the hidden layers. In this manner, the inputs are the outputs of activation functions in where the inputs are multiplied by the connection weights. This activation function (neuron transfer function) determines the output. In practice, one single neuron is not capable of handling problems. That's why networks composed of neurons are being used. In Figure \ref{fig:2}, the architecture of a multilayer perceptron is shown.
In this study, we consider a feed forward neural network with one input layer, one hidden layer and one output layer. In feed forward neural networks, the information moves in only one direction, from input nodes to the hidden and output nodes. Fig. \ref{fig:1} is an example of feed forward neural network.

Solving differential and eigenvalue equations via trained ANN have some major advantages compared to standard numerical techniques \cite{41}:

\begin{itemize}
\item A compact form of the solution is derivable since finite differences are not utilized,
\item Increasing number of sampling points does not increase the computational complexity very quickly. \end{itemize}

These features are important for solving few body systems by Schrödinger equation. Solving a two body system like mesons by Schrödinger equation seems quite easy rather than three-body systems like baryons. For example to describe a baryon system as a bound state of three quarks is complicated. The configuration of particles can be done via a coordinate transformation called Jacobi coordinates. Faddeev equations are also being used for quantum three body problems \cite{42}.

\subsection{Mathematical Model of an Artificial Neural Network}
The general method for solving differential equations is presented in \cite{43} with the quantum mechanical applications in \cite{44,45,46,47}. The relationship of the input-output of the layers can be written as follows:
\begin{eqnarray}
o_i&=&\sigma(n_i), \\
o_j&=&\sigma(n_j), \\
o_k&=&\sigma(n_k),
\end{eqnarray}
where $i$ is for input, $j$ is for hidden and $k$ is for output layers. Input to the perceptrons are given as
\begin{eqnarray}
n_i&=&(\text{Input signal to the NN}), \\
n_j&=& \sum_{i=1}^{N_i} \omega_{ij}o_i+ \theta_j, \\
n_k&=& \sum_{i=1}^{N_j} \omega_{jk}o_j+ \theta_k,
\end{eqnarray}
where, $N_i$ and $N_j$ represents the numbers of the units belonging to input and hidden layers, $\omega_{ij}$ is the synaptic weight parameter which connects the neurons $i$ and $j$ and $\theta_j$ represents threshold parameter for the neuron $j$ and $\theta_j$ is the threshold parameter \cite{47}. The output of the network can be written as
\begin{equation}
o_k=\sum_{j=1}^{b_n} \omega_{jk}\sigma \left( \sum_{i=1}^{a_n}\omega_{ij}n_i +\theta_j \right)+ \theta_k. \label{eqn1}
\end{equation}
Derivative of this function is needed in further evaluation of error function. This can be obtained as
\begin{eqnarray}
\frac{\partial o_k}{\partial \omega_{ij} }&=& \omega_{jk} \sigma^{(1)}(n_j)n_i,\\
\frac{\partial o_k}{\partial \omega_{jk} }&=& \sigma(n_j)\delta_{kk^\prime},\\
\frac{\partial o_k}{\partial \theta_j }&=& \omega_{jk} \sigma^{(1)}(n_j),\\
\frac{\partial o_k}{\partial \theta_{k^\prime} }&=&\delta_{kk^\prime}.
\end{eqnarray}

In this work we use a sigmoid function
\begin{equation}
\sigma(z)=\frac{1}{1+e^{-z}} \label{sigmoid}
\end{equation}
as an activation function since it is possible to derive all the derivatives of $\sigma(z)$ in terms of itself. This differentiability is an important aspect for the Schrödinger equation.

\subsection{Application to Quantum Mechanics}
Following the work of \cite{44}, let us consider the differential equation:
\begin{equation}
H\Psi(r)=f(r) \label{eqn2}
\end{equation}
where $H$ is a linear operator, $f(r)$ is a known function and  $\Psi(r)=0$ at the boundaries. In order to solve this differential equation, a trial function 
\begin{equation}
\Psi_t(\textbf{r})=A(\textbf{r})+B(\textbf{r}, \textbf{$\lambda$})N(\textbf{r}, \textbf{p})
\end{equation}
of the form can be written which uses a feed forward neural network with parameter vector $\textbf{p}$ and $\textbf{$\lambda$}$ to be adjusted. The parameter $\textbf{p}$ refers to the weights and biases of the neural network. The functions $A(\textbf{r})$ and $B(\textbf{r}, \textbf{$\lambda$})$ should be specified in a convenient way so that $\Psi_t(\textbf{r})$ satisfies the boundary conditions regardless of the $\textbf{p}$ and $\textbf{$\lambda$}$ values.  To obtain a solution for Eqn. (\ref{eqn2}), the collocation method can be used and the differential equation can be transformed into a minimization problem 
\begin{equation}
\underset{p,\lambda}{\min} \sum_i \left[ H\Psi_t(r_i)-f(r_i)  \right]^2. \label{dif2}
\end{equation}
For Schrödinger equation Eqn. (\ref{eqn2}) takes the form
\begin{equation}
H\Psi(r)=\epsilon \Psi(r) 
\end{equation}
with the boundary condition, $\Psi(r)=0$. In this case, the trial solution can be written as
\begin{equation}
\Psi_t(r)=B(\textbf{r}, \textbf{$\lambda$})N(\textbf{r}, \textbf{p})
\end{equation}
where $B(\textbf{r}, \textbf{$\lambda$}) =0$ at boundary conditions for a range of $\lambda$ values. By discretizing the domain of the problem, it is transformed into a minimization problem with respect to the parameters $\textbf{p}$ and $\textbf{$\lambda$}$
\begin{equation}
E(\textbf{p},\textbf{$\lambda$})=\frac{\sum_i \left[ H\Psi_t(r_i, \textbf{p},\textbf{$\lambda$})-\epsilon \Psi_t(r_i, \textbf{p},\textbf{$\lambda$})  \right]^2}{\int \vert \Psi_t \vert^2 d\textbf{r}}
\end{equation}
where $E$ is the error function and $\epsilon$ can be computed as
\begin{equation}
\epsilon=\frac{\int  \Psi_t^{\ast} H \Psi_t  d\textbf{r}}{\int \vert \Psi_t \vert^2 d\textbf{r}}.
\end{equation}

\section{\label{sec:level3}Numerical Results and Discussion}
We consider the Hamiltonian which was formulated by Semay and Silvestre-Brac in \cite{48} to study tetraquark systems. The Hamiltonian reads as follows
\begin{equation}
H=\sum_i \left( m_i+\frac{\bf{p}_i^2}{2m_i} \right)-\frac{3}{16}\sum_{i<j} \tilde{\lambda}_i \tilde{\lambda}_j v_{ij}(r_{ij})
\end{equation}
with the potential 
\begin{eqnarray}
v_{ij}(r)&=&-\frac{\kappa(1-e^{-\frac{r}{r_c}})}{r}+\lambda r^p \\ \nonumber &+& \Lambda +
           \frac{2\pi}{3m_im_j}\kappa^\prime (1-e^{-\frac{r}{r_c}})\frac{e^{-\frac{r^2}{r_0^2}}}{\pi^{3/2}r_0^3} \bf{\sigma_i} \bf{\sigma_j}, \label{potential}
\end{eqnarray}
where $r_0(m_i,m_j)=A \left(\frac{2m_im_j}{m_i+m_j} \right)^{-B}$, $A$ and $B$ are constant parameters, $\kappa$ and $\kappa^\prime$ are parameters, $r_{ij}$ is the interquark distance $\vert \bf{r_i}-\bf{r_j \vert}$, $\sigma_i$ are the Pauli matrices and $\tilde{\lambda}_i$ are Gell-Mann matrices. We used the potential called AL1 with the parameters given as \citep{48}

\begin{eqnarray}
m_u=m_d&=& 0.315 ~ \text{GeV}, \nonumber  \\ 
m_s&=&0.577 ~ \text{GeV},  \nonumber \\ 
m_b &=& 5.227 ~\text{GeV}, \nonumber \\ 
m_c &=& 1.836 ~\text{GeV}, \nonumber \\ 
\kappa &=& 0.5069, \nonumber \\ 
\kappa^\prime &=& 1.8609, \nonumber \\ 
\lambda &=& 0.1653~ \text{GeV}^2, \nonumber \\ 
\Lambda &=& -0.8321~ \text{GeV},  \nonumber \\
B &=& 0.2204, \nonumber \\
A &=& 1.6553~ \text{GeV}^{B-1}, \nonumber \\
r_c &=& 0,
\end{eqnarray}

and the masses as \cite{49}

\begin{eqnarray*}
B_s^0 &=& 5366  ~ \text{MeV} \\
\pi &=& 139 ~ \text{MeV} \\
B_s^* &=& 5415 ~ \text{MeV} \\
\rho &=& 770 ~ \text{MeV} \\
B^+ &=& 5279 ~ \text{MeV} \\
\bar{K}^0 &=& 497 ~ \text{MeV} \\
B^{*+} &=& 5325 ~ \text{MeV} \\
\bar{K}^{*0} &=& 892 ~ \text{MeV} \\
D_s^- &=& 1968 ~ \text{MeV} \\
D^0 &=& 1864 ~ \text{MeV}.
\end{eqnarray*}

Constructing wave function for four-body system is straightforward. We have used a wave function which was proposed in \cite{50}

\begin{eqnarray}
\psi_r(\text{1234}; \bf{x_1},\bf{x_2},\bf{x_3})&=& C_i(1234)T_j(1234)S_k(1234)\nonumber \\ &\times & E_p(\bf{x_1},\bf{x_2},\bf{x_3}),
\end{eqnarray}
where 1,2 and 3,4 denotes the quarks and antiquarks, respectively, $C_i$ is the color part, $T_j$ is the isospin, $S_k$ is the spin and $E_p$ is the spatial parts, respectively. The spatial basis states are composed of harmonic oscillator wave functions. For this, three Jacobi coordinates $\bf{x_1}$ (diquark extension), $\bf{x_2}$ (antidiquark extension) and $\bf{x_3}$ (diquark-antidiquark distance) are defined as \cite{50}
\begin{eqnarray}
b\bf{x_1}&=&\left[ \frac{2 \omega_1 \omega_2}{\omega_{12}} \right]^{1/2} (\bf{r_1}-\bf{r_2}),  \nonumber \\
b\bf{x_2}&=&\left[ \frac{2 \omega_3 \omega_4}{\omega_{34}} \right]^{1/2} (\bf{r_3}-\bf{r_4}), \nonumber \\ 
b\bf{x_3}&=&\left[ \frac{2 }{\omega \omega_{12} \omega_{34}} \right]^{1/2} \nonumber \\  & \times &  \left[ \omega_{34}  (\omega_1 \bf{r_1}+ \omega_2 \bf{r_2}) - \omega_{12} (\omega_3 \bf{r_3}+ \omega_4 \bf{r_4}) \right].
\end{eqnarray}
In these equations, a reference length $b$ is chosen arbitrarily to make sure that Jacobi coordinates $\bf{x_i}$ are dimensionless. Similarly, a reference mass $m$ is chosen and $\omega_i=m_i/m$ are dimensionless parameters proportional to the actual masses. The definitions $w_{ij}=\omega_i + \omega_j$ and $\omega=w_{12}+w_{34}=\sum_i \omega_i=M/m$ where $M$ is the total mass of the particles are used for compactness.

\begin{equation}
\psi(r)=e^{-\beta r^2}  N(x_i,u,w,v) \psi_r(\text{1234}; \bf{x_i}), ~ \beta \geq 0
\end{equation}
with $N$ being a feed forward neural network with one hidden layer and $m$ sigmoid hidden units
\begin{equation}
N(x_i,u,w,v)=\sum_{j=1}^m v_j \sigma(w_j x_i+u_j).
\end{equation}

By employing this approach it is possible to obtain energy eigenvalues of the Schrödinger equation. We trained the network with 200 equidistance points in the intervals $0 < r < 1$ with $m=8$ and solved the Schrödinger equation with $\psi(0)=0$ at the boundaries.  Table \ref{tab:table1} gives the mass values for $X(5568)$ and Table \ref{tab:table2} for $X_c$.

\begin{table}[H]
\caption{\label{tab:table1}Mass values of $X(5568)$. Results are in MeV.}
\begin{ruledtabular}
\begin{tabular}{ccccccc}
 $X(5568)$ & This work &  \cite{51} &  \cite{52} & \cite{28} & \cite{21}
  \\
\hline
$su\bar{d}\bar{b}$ & 5885 & $5864 \pm 158 $ & $5584 \pm 137 $  \\
$B_s^0 \pi $ & 5743 &  & & 5507 &  \\
$ B_s^* \rho  $ & 6186 &  & & 6182 &  \\
$ B^+ \bar{K}^0 $ & 5845 &  & & 5774 & $5757 \pm 145 $   \\
$ B^{*+} \bar{K}^{*0} $ & 6207 &  & & 6233 & &  \\

\end{tabular}
\end{ruledtabular}
\end{table}

\begin{table}[H]
\caption{\label{tab:table2} Mass values of $X_c$. Results are in MeV except Ref. \cite{27} which is in GeV.}
\begin{ruledtabular}
\begin{tabular}{ccccccc}
  & $X(5568)$ &  \cite{33} &  \cite{33} & \cite{27}
  \\
\hline
Mass & 2480 & $2590 \pm 60 $ & $2634 \pm 62 $ & $2.55 \pm 0.09 $  \\

\end{tabular}
\end{ruledtabular}
\end{table}
 
According to our framework, $X(5568)$ is light for an $su\bar{b}\bar{d}$ tetraquark or molecular state. The mass difference is at the order of  $\approx 300 ~ \text{MeV}$. Taking into account that the $\Xi_b$ baryon which has $usb$ quark structure has a mass of 5797 MeV, it would be a puzzling situation if  $su\bar{d}\bar{b}$ structure have a lower mass with an additional quark than $\Xi_b$ baryon.

\section{\label{sec:level4} Summary and Concluding Remarks}
In this work, we obtained mass of $X(5568)$ resonance and hypothetical partner $X_c$. The prominent feature of this resonance is that it is the first state that contains four different flavors of quark. Although the other collaborations  LHCb, CMS, ATLAS and CDF have not confirmed the existence of this state up to now, the statistical significance of 5.1 $\sigma$ in the $B_s^0 \pi^{\pm}$ invariant-mass spectrum challenges our understanding of the quark model. In the original quark model framework such exotic states or better to say multi-quark states were predicted by Gell-Mann. Therefore it should be not surprising the existence of four-quark state with all different flavors.

Due to some advantages provided by artificial neural networks such as continuity of solution over all the domain of integration and not increasing of computational complexity when the sampling points and number of dimensions involved, such elaborations can be made more safely. Neural networks are also being used in quantum information theory \cite{53,54}.

As can be seen from Table \ref{tab:table1}, all the mass values (except for the last case) are above their relative thresholds. In this view, they are not meson-meson bound states but resonances. According to our framework, $X(5568)$ cannot be explained as a molecular or a diquark-antidiquark resonance. 

The mass value calculation alone itself does not corroborate the internal structure of any state, exotic or not but gives an idea about the validation of the models. Further theoretical and experimental studies would clarify the status of $X(5568)$ and also $X_c$.

\end{document}